\documentclass[submission, Phys]{SciPost}

\hypersetup{
    colorlinks,
    linkcolor={red!50!black},
    citecolor={blue!50!black},
    urlcolor={blue!80!black}
}

\usepackage[bitstream-charter]{mathdesign}
\usepackage{bm,amsmath,verbatim}
\DeclareSymbolFont{usualmathcal}{OMS}{cmsy}{m}{n}
\DeclareSymbolFontAlphabet{\mathcal}{usualmathcal}

\begin{document}

\begin{center}{\Large \textbf{
Three-Dimensional Quantum Hall Effect in Topological Amorphous Metals\\
}}\end{center}

\begin{center}
Jiong-Hao Wang\textsuperscript{1} and
Yong Xu\textsuperscript{1,2$\star$}
\end{center}

\begin{center}
{\bf 1} Center for Quantum Information, IIIS, Tsinghua University, Beijing 100084, People's Republic of China
\\
{\bf 2} Hefei National Laboratory, Hefei 230088, People's Republic of China
\\
${}^\star$ {\small \sf yongxuphy@tsinghua.edu.cn}
\end{center}


\section*{Abstract}
{\bf
Weyl semimetals have been theoretically predicted to become topological metals with 
anomalous Hall conductivity in amorphous systems. 
However, measuring the anomalous Hall conductivity in realistic materials, particularly 
those with multiple pairs of Weyl points, is a significant challenge.
If a system respects time-reversal symmetry, then the anomalous Hall conductivity even vanishes. 
As such, it remains an open question how to probe the Weyl band like topology in amorphous 
materials.
Here, we theoretically demonstrate that, under magnetic fields, a topological metal slab in amorphous systems 
exhibits three-dimensional quantum Hall effect, even in time-reversal invariant systems, thereby 
providing a feasible approach to exploring Weyl band like topology in amorphous materials.
We unveil the topological origin of the quantized Hall conductance by calculating the Bott index. 
The index is carried by broadened Landau levels with bulk states spatially localized except at critical transition energies.
The topological property also results in edge states localized at distinct hinges on two opposite surfaces.
}

\vspace{10pt}
\noindent\rule{\textwidth}{1pt}
\tableofcontents\thispagestyle{fancy}
\noindent\rule{\textwidth}{1pt}
\vspace{10pt}

\section{Introduction}
\label{sec:intro}
Weyl semimetals~\cite{Burkov2016NM,Hasan2016NM,Vishwanath2018RMP,Xu2019FP,Ding2021RMP}, 
a celebrated example of gapless topological phases, have attracted considerable 
attention due to both fundamental interest in the emergent Weyl femions as well as novel topological properties
and potential applications in future electronic
devices~\cite{Savrasov2011PRB,Ran2011PRB,Balents2011PRL,Fang2011PRL,Brouwer2014PRL,
	Vishvanath2014NC,Buljan2015PRL,Udagawa2015PRL, Dai2015PRX,Zhang2015PRL,Xie2015PRL,
	Hasan2015Science,Ding2015PRX,Soljacic2015Science,
	ZhangC2015PRL,Bernevig2015Nature,Vishwanath2016SciRep,Sarma2016PRX,Kaminiski2016NM,
	Zhou2016NP,Nagaosa2016PRL,Duan2016PRA,Gop2018PRL,Rad2018ARCMP,Xie2019PRB,
	Chen2019Science,Beiden2019Science,Jiang2021PRB,Zhu2023arXiv}.
Their energy bands touch linearly at a Weyl point protected by the Chern number over a two-dimensional surface 
enclosing the Weyl point in three-dimensional (3D) momentum space. 
By bulk-boundary correspondence, topological properties in Weyl semimetals manifest in the surface Fermi arcs 
connecting Weyl points with opposite chirality, which can give rise to 
\textit{anomalous} Hall effect~\cite{Ran2011PRB,Balents2011PRL}.
Remarkably, such Fermi arc states assisted by bulk chiral Landau levels in Weyl semimetals under magnetic fields
can result in 3D quantum Hall effect~\cite{Xie2017PRL,Xiu2017NC2,Kawasaki2017NC,
	Stemmer2018PRL,Xiu2019Nature,
	Xie2020PRL,Xiu2021NRP,Xie2021PRL,Xie2021npj,Sheng2022CPB,Nag2022PRB,JH2025PRB,JH2025CFL}.

Apart from crystalline materials with translational symmetry, nature also provides us with abundant amorphous solids 
without translational symmetry~\cite{Zallen1997book}. 
Although topological physics is established in crystalline materials, 
recent studies have suggested that amorphous systems can also display various topological
phases~\cite{Shenoy2017PRL,Chong2017PRB,Irvine2018NP,Prodan2018JPA,Ojanen2018NC,
	Zhang2019PRB,Chern2019EPL,Xu2019PRL,Fazzio2019NanoLett,Ojanen2020PRR,Bhatta2020PRB, Roy2020PRR,
	Liu2020Research,Zhang2020LSA,Grushin2020PNAS,Ojanen2020PRR2,Lew2021Mater,Xu2021PRL,Xu2021PRL2,Huang2022PRL,
	Grushin2022book,Kvorning2022PRL,Chowdhury2023PRL,Hellman2023NM,Ojanen2023PRR,Fleury2023SA,Grushin2023EPL,Xu2023arxiv,YangReview2023}
	and . 
In particular, it has been theoretically shown that Weyl semimetals without time-reversal symmetry (TRS) 
will become topological metals 
with nonzero anomalous Hall conductivity when lattice sites are positioned
randomly~\cite{Xu2019PRL}.
However, it is significantly challenging to experimentally measure the anomalous Hall conductivity in realistic materials,
especially for systems with multiple pairs of Weyl points,
whose topology can be characterized by the spectral localizer~\cite{Hermann2022EPL,Alexander2022PRB,Hermann2023JMP,Kahlil2023PRL,Franca2024PRM}. 
If a system respects TRS, then the anomalous Hall conductivity even 
vanishes in general.
In addition, conventional methods to probe Weyl points and surface states using angle-resolved photoemission spectroscopy 
(ARPES)~\cite{Hasan2015Science,Ding2015PRX}  may be difficult due to the absence of a well-defined momentum space [see Appendix \ref{AppB} for the numerically simulated ARPES spectra]. 
We therefore ask whether 3D quantum Hall effect can exist in amorphous Weyl materials under magnetic fields,
which may provide a feasible approach to probing Weyl band like topology in amorphous materials.                

\begin{figure}[t]
	\centering
	\includegraphics[width=0.8\textwidth]{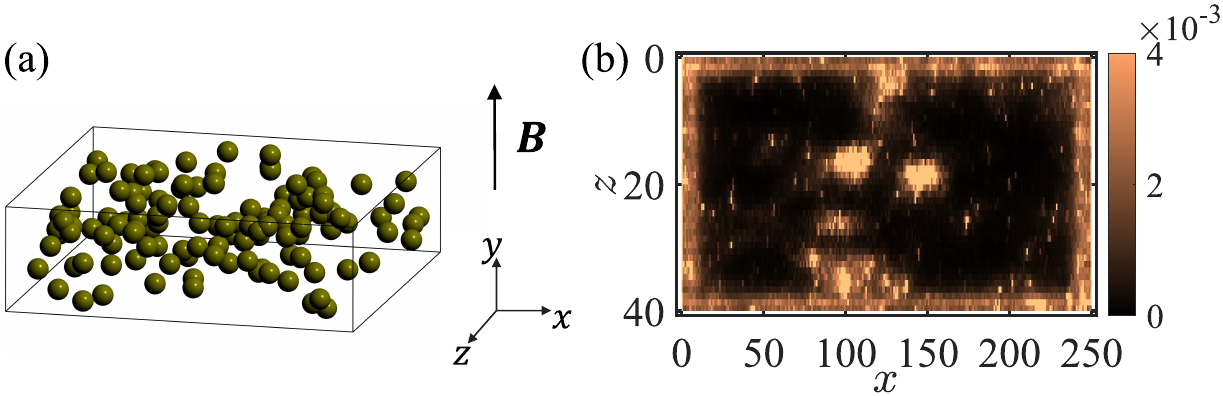}
	\caption{
		(a) Schematic of a random site configuration in a slab geometry with magnetic fields ${\bm B}$ along the $y$ direction.
		(b) Configuration averaged local density of states (DOS) summed over coordinates along $y$ 
		for the Hamiltonian (\ref{Ham1}) on amorphous lattices with system size $L_x=250$, $L_y=20$, and $L_z=40$ 
		subjected to magnetic fields, illustrating the presence of edge states.
	}
	\label{Fig1}
\end{figure}

In this work, we theoretically demonstrate that 3D quantum Hall effect can occur in amorphous systems 
by studying a tight-binding model on a random lattice under a magnetic field [see Fig.~\ref{Fig1}(a)].
We find the existence of quantized plateaus of the Hall conductance contributed by edge states
[see Fig.~\ref{Fig1}(b)].
The edge states have a topological origin that can be characterized by the Bott index.
Despite the presence of strong structural disorder, we show that the bulk states can still form Landau
levels, albeit broadened. However, we show that these levels are localized, except at 
the critical transition point between two plateaus. As a consequence, only edge states contribute to the transport,
ultimately resulting in the quantized Hall conductance.
In a regular lattice, the 3D quantum Hall effect is explained 
through semiclassical analysis based on lattice momentum ${\bm k}$ by showing that
electrons cycle along the Fermi arcs on top and bottom surfaces aided by chiral Landau levels tunneling between the opposite 
surfaces~\cite{Xie2017PRL,Xie2020PRL}.
In amorphous systems, while the translational symmetry is entirely lost so that the ${\bm k}$ space description 
breaks down, we still observe the edge states localized at distinct hinges on opposite surfaces.
We construct a time-reversal invariant model with double pairs of Weyl points when lattice sites are positioned 
regularly. In both regular and amorphous geometries, it has no \textit{anomalous} Hall conductivity.
However, when a magnetic field is applied, we see the appearance of nonzero quantized Hall conductance, 
demonstrating that 3D quantum Hall effect can be used as an indicator of topology for amorphous Weyl materials with vanishing 
\textit{anomalous} Hall conductivity.
Finally, we show the three-dimensional quantum Hall effect can also exist under the tilted magnetic field.
In this case, we find a region for the time-reversal symmetric model where the quantized Hall conductance is 
		present on amorphous lattices but absent on the regular lattice.
		The phenomenon is attributed to the additional pair of counter-propagating edge channels on each edge, 
		which are localized on amorphous lattices.

\section{Model Hamiltonian}
To demonstrate the existence of 3D quantum Hall effect in amorphous systems, 
we consider the following tight-binding Hamiltonian 
\begin{equation}\label{Ham1}
	\hat{H}=\sum_{\bm r} [\hat{c}_{\bm r}^\dagger T_0\hat{c}_{\bm r}+\sum_{\bm d} t(|{\bm d}|)
	\hat{c}_{\bm r+\bm d}^\dagger 
	T(\hat{{\bm d}})\hat{c}_{\bm r}	],
\end{equation} 
where  $\hat{c}_{\bm r}^\dagger=(\hat{c}_{{\bm r},\uparrow}^\dagger, \hat{c}_{{\bm r},\downarrow}^\dagger)$ 
with 
$\hat{c}_{{\bm  r},\sigma}^\dagger$ being the fermionic creation operator of spin $\sigma$ at
position ${\bm r}$, and $\hat{c}_{\bm r}$ being its Hermitian conjugate, the corresponding 
annihilation operator.
$T_0=M(k_w^2-6) \\ \sigma_z  +(4D_2+2D_1)\sigma_0$ is the onsite term, and
$T(\hat{{\bm d}})=M\sigma_z+i(\gamma/{2}) (\hat{d}_x\sigma_x+\hat{d}_y\sigma_y)-(D_2\hat{d}_x^2+D_1\hat{d}_y^2+\\ D_2\hat{d}_z^2)\sigma_0$
is the hopping matrix from the site at
position ${\bm r}$ to a different site at ${\bm r+\bm d}$, where $\sigma_0$ is the identity matrix,
$\sigma_{\nu}$ ($\nu=x,y,z$) are Pauli matrices, and $\hat{{\bm d}}={{\bm d}}/{|{\bm d}|}=(\hat{d}_x,\hat{d}_y,\hat{d}_z)$
is the normalized separation vector.
In the amorphous case, we set $t(|{\bm d}|)=\Theta(R_c-|{\bm d}|)e^{-\lambda(|{\bm d}|/a-1)}$,
decaying exponentially to simulate real materials. Here, $\Theta(R_c-|{\bm d}|)$ is the Heaviside 
step function such that
hoppings for $|{\bm d}|>R_c$ are cut off and the lattice constant is set to $a=1$. 
In a regular lattice with only nearest-neighbor hopping, 
i.e. the hopping strength $t(|{\bm d}|)$ set to be $1$ for nearest neighbor and $0$ otherwise, 
the Hamiltonian (\ref{Ham1}) reduces to a paradigmatic Weyl
semimetal model corresponding to a momentum space Hamiltonian $H(\bm{k})=
M(k_w^2-\bm{k}^2)\sigma_z+\gamma(k_x\sigma_x+k_y \\ \sigma_y)+D_1k_y^2+D_2(k_x^2+k_z^2)$ 
in the continuum limit, where $M$, $k_w$, $\gamma$, $D_1$ and $D_2$ are real system parameters.
The Weyl points are located at ${\bm k}=(0,0,\pm k_w)$ with energy
$E_w=D_2k_w^2$.
When $D_1, D_2\neq 0$, Fermi arcs are bent, which is essential to 3D quantum Hall effect when the Fermi energy lies at 
the energy of Weyl points~\cite{Xie2017PRL,Xie2020PRL}.

\begin{figure}[t]
	\centering
	\includegraphics[width=0.65\textwidth]{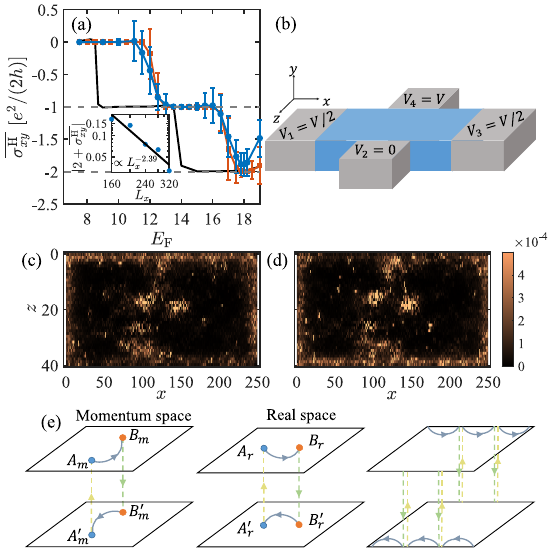}
	\caption{
		(a) Sample averaged Hall conductance $\overline{\sigma_{xy}^{\text{H}}}$ (blue line) and Bott 
		index (red line) versus the Fermi energy $E_\text{F}$ of the Hamiltonian (\ref{Ham1}) under
		magnetic fields on amorphous lattices compared with the Hall conductance on regular lattices (black line). 
		The system size for the Hall conductance is $L_x=200,L_y=20,L_z=40$ and that for the Bott index is 
		$L_x=25,L_y=20,L_z=44$.
		Inset: the Hall conductance at $E_\text{F}=18.2$ for $L_x=160,200,240,280,320$ (blue points) fitted by a 
		black line in the logarithmic scale, exhibiting the power-law decay. 
		We keep the ratio $L_x/L_z=5$, and $L_y=20$ is fixed.	
		(b) Schematic illustration of a four-terminal setup to extract the Hall conductance.
		(c-d) Sample averaged local DOS on the top and bottom layer, respectively,
		calculated for $L_x=250,L_y=20,L_z=40$
		within $L_y-1 \leq z \leq L_y$ and $0 \leq z \leq 1$
		(e) Schematic of cyclic trajectories of an electron under magnetic fields [in momentum space (left 
		panel) and real space (middle panel)] and the resulting skipping edge states (right panel). 
	}
	\label{Fig2}
\end{figure}

In the amorphous case, we consider sites that are positioned completely randomly in a slab, which is thin along $y$, necessary to the observation of quantum Hall effect even in the pristine regular topological semimetal [see Appendix \ref{AppC}].
To be explicit, the positions of lattice sites are sampled from uncorrelated uniform distributions
within the interval $[0, L_\nu)$, where $L_\nu$ is the length of the system along $\nu$ with $\nu=x,y,z$.
We take the total number of sites as $N=\rho V$ with the volume of the system $V=L_x L_y L_z$ and 
the average density $\rho=1$.   
An external magnetic field ${\bm B}=(0,B,0)$ is applied in the $y$ direction [see Fig.~\ref{Fig1}(a)].
The magnetic field modifies the hopping matrix $T(\hat{{\bm d}})$ by a phase 
factor $e^{-i\frac{e}{\hbar}\int_{\bm r}^{\bm {r}+\bm {d}}{\bm A}\cdot d{\bm r}^{\prime} }$ where the 
physical constants are set to be $e=\hbar=1$ and we choose a gauge such that the vector potential is ${\bm A}=(Bz,0,0)$.  
Note that without changing the physics qualitatively, we follow previous works to ignore 
the contribution of the Zeeman effect~\cite{Xie2017PRL,Xie2020PRL}.
In numerical computations throughout the paper, we take $M=8, {\color{red}\gamma}=50, k_w=1.5, D_1=1, D_2=4, R_c=2.5, B=\pi/22$ and $\lambda=3$ 
without loss of generality, and the results are averaged
over 100 sample configurations unless stated otherwise.  

In a regular lattice, the Hamiltonian (\ref{Ham1}) hosts Fermi arcs on top and bottom surfaces viewed in the $y$ direction.
The Fermi arcs contribute to edge channels with the assistance of bulk chiral Landau levels under magnetic fields, 
giving rise to 3D quantum Hall effect~\cite{Xie2017PRL, Xie2020PRL}. 
In the amorphous case, while previous study suggests the existence of surface states despite the 
absence of translational symmetry~\cite{Xu2019PRL}, it is unclear whether isolated Landau levels can survive so as to 
result in 3D quantum Hall effect. To investigate the possibility, we will calculate the Hall conductance, 
the Bott index, the DOS and the localization properties of the Hamiltonian (\ref{Ham1}) in random site configurations.

\section{Hall conductance, Bott index and edge states}
To show the presence of 3D quantum Hall effect 
in amorphous systems, we calculate the Hall conductance using Landauer-B$\ddot{\mbox{u}}$ttiker 
formula under four-terminal measurement with voltage $V_1=V_3=V/2, V_4=V$ and $V_2=0$ 
applied on ideal leads, as shown in Fig.~\ref{Fig2}(b). 
Note that the terminal configuration here is different from that for measuring anomalous Hall effect, e.g., in Ref.~\cite{Xu2019PRL}, with the surfaces connected to terminal 2 and 4 carrying no Fermi arcs.
The Hall conductance $\sigma_{xy}^{\text{H}}$ is given by~\cite{Xu2019PRL}
\begin{equation}\label{Hall1}
	\sigma_{xy}^{\text{H}}=\frac{e^2}{2h}(T_{14}-T_{12}),	
\end{equation}
where $T_{ij}$ is the transmission probability from lead $j$ to $i$ computed numerically using
non-equilibrium Green's function 
method~\cite{Mackinnon1985ZPB, Mackinnon1985ZPB2, Datta1997, Wang2007PRB}. 
In the phase with 3D quantum Hall effect, $\sigma_{xy}^{\text{H}}$ is quantized to an integer multiple of $e^2/(2h)$.   
Note that in the quantized region, the Hall conductance determined by Eq. (\ref{Hall1}) is 
identical to that measured under constant longitudinal and no transverse current up to a multiplier 
$1/2$~\cite{Xie2017PRL,Xie2020PRL}. 

In Fig.~\ref{Fig2}(a), we plot the sample averaged Hall conductance $\overline{\sigma_{xy}^{\text{H}}}$ 
with respect to the Fermi energy $E_\text{F}$ (blue line). Remarkably, we see a plateau quantized at
$\overline{\sigma_{xy}^{\text{H}}}=-e^2/(2h)$ in the region $13 \lesssim E_\text{F} \lesssim 16$, establishing the presence of
amorphous 3D quantum Hall effect. 
Specifically, as $E_\text{F}$ is increased from $E_\text{F}=7.5$, we see $\overline{\sigma_{xy}^{\text{H}}}=0$ 
for $E_\text{F} \lesssim 11$ and
then $\overline{\sigma_{xy}^{\text{H}}}$ undergoes a transition around $E_\text{F} \approx 12$ into the plateau at 
$\overline{\sigma_{xy}^{\text{H}}}=-e^2/(2h)$.
As we further raise $E_\text{F}$, $\overline{\sigma_{xy}^{\text{H}}}$ declines and reaches a plateau near 
$\overline{\sigma_{xy}^{\text{H}}}=-2e^2/(2h)$ for $17.8 \lesssim E_\text{F} \lesssim 18.2$.
Our numerical calculations further show that $|2+\overline{\sigma_{xy}^{\text{H}}}| \propto L_x^{-2.39} $ with
system size $L_x$ at $E_\text{F}=18.2$
[see the inset of Fig.~\ref{Fig2}(a)], suggesting that within this region, $\overline{\sigma_{xy}^{\text{H}}}$ can 
reach the quantized value of $-2e^2/(2h)$ in the thermodynamic limit. 
Compared with the Hall conductance $\overline{\sigma_{xy}^{\text{H}}}$ of a system with regular lattice sites (black line), 
we see that the corresponding plateaus for the amorphous system move toward higher $E_\text{F}$,
which is attributed to the shift of Landau levels in the amorphous case as will be discussed later. 

To reveal the topological nature of 3D quantum Hall effect, we calculate the Bott 
index~\cite{Loring2010JMP,Loring2010EPL} defined as
\begin{equation}\label{Bott}
	\mathrm{Bott}=\frac{1}{2 \pi}\mathrm{Im Tr} \log (U_z U_x U_z^\dagger U_x^\dagger),	
\end{equation}
where $[U_\nu]_{mn}=\langle \psi_m|e^{i2\pi \hat{\nu}/L_\nu}|\psi_n\rangle$ ($\nu=z,x$) 
with $\hat{\nu}$ being position operators along $\nu$ and $|\psi_n\rangle$ 
$(1,2, ..., N_{occ})$ being occupied single-particle states under $E_\text{F}$.
In our numerical computations, we impose periodic boundary conditions (PBCs) along 
$x$ and $z$ and open boundary conditions (OBCs) along $y$, 
given the fact that the surface states on top and bottom surfaces are crucial to the 3D quantum 
Hall effect.

Figure~\ref{Fig2}(a) shows the computed Bott index (red line), which agrees well with 
the Hall conductance (blue line). In fact, the Bott index is less affected by finite-size effects due to 
the imposed PBCs. To be specific, for $17.8 \lesssim E_\text{F} \lesssim 18.2$, the Bott index is well quantized at
$\mathrm{Bott}=-2$, further verifying the existence of the plateau with $\overline{\sigma_{xy}^\text{H}}=-2e^2/(2h)$
in thermodynamic limit.
In addition, within the transition region between quantized plateaus, the Bott index exhibits a sharper decline 
than the Hall conductance.
The consistence between the Bott index and the Hall conductance thus establishes the topological origin of the 3D 
quantum Hall effect.

Because of the topological properties, we expect the existence of hinge states on top and bottom surfaces, leading
to the 3D quantum Hall effect. To identify the hinge states, we calculate the local DOS at energy $E$ and position ${\bm r}$ 
based on
\begin{equation}\label{LDOS}
	\rho({\bm r},E )=-\frac{1}{\pi}\mathrm{Im Tr} \, G_{\bm r \bm r}(E)
\end{equation}
using the recursive Green's function method~\cite{Mackinnon1985ZPB, Mackinnon1985ZPB2}.
Here, $G_{\bm r\bm r}(E)=\langle {\bm r}|(E-H+i0^+)^{-1}|{\bm r} \rangle$ where 
$H$ is the single-particle first quantization form of the Hamiltonian (\ref{Ham1}) under OBCs along all three directions,
$0^+$ denotes a positive infinitesimal, and $|{\bm r} \rangle$ is the site state at position ${\bm r}$.
Figure~\ref{Fig1}(b) shows the local DOS at $E=14$ summed over the coordinates along $y$, remarkably
exhibiting the existence of edge states despite the presence of strong structural disorder. 
Note that the small bright regions in the interior are contributed by localized bulk states (here the bulk
states refer to the states calculated under OBCs along $y$ and PBCs along $x$ and $z$ 
and edge states can only be obtained by further applying OBCs along either $x$ or $z$ or both) 
due to the finite number of samples.  
Ref.~\cite{Xu2019PRL} shows that a Weyl semimetal develops into a topological metal in a random lattice with surface states.
Such edge states on top and bottom surfaces thus arise from these surface states under magnetic fields. 

Figure~\ref{Fig2}(c) and (d) further display the local DOS at the top and bottom layer along $y$, respectively.
It surprisingly illustrates that the hinge states are primarily concentrated at the $z=0$ edge on the top surface 
[Fig.~\ref{Fig2}(c)] while at the other edge on the bottom surface [Fig.~\ref{Fig2}(d)]. 
In a regular lattice, such a distribution of hinge states on opposite surfaces arises from Fermi arc states 
assisted by bulk chiral Landau levels ~\cite{Xie2017PRL,Xie2020PRL}.
Specifically, 
according to the semiclassical equations of motion, $\dot{\bm r}=\partial \varepsilon _{\bm k}/(\hbar \partial {\bm k})$ and 
$\hbar \dot{\bm k}=-e\dot{\bm r}\times {\bm B}$~\cite{Niu2010RMP},
if an electron starts at $A_m$ in momentum space and $A_r$ in real space, then under magnetic fields, 
it will move from $A_m$ to $B_m$ in momentum space following the Fermi arc. Meanwhile, the electron
moves from $A_r$ to $B_r$ in real space [see Fig.~\ref{Fig2}(e)].
At $B_m$, the electron tunnels to $B_m^\prime $ on the opposite surface through a bulk chiral Landau level.
In real space, the electron tunnels to $B_r^\prime$ on the opposite surface. 
On the bottom surface, the electron further moves from $B_m^\prime$ ($B_r^\prime$) to $A_m^\prime$ 
($A_r^\prime$) in momentum (real) space, completing cyclic motions in both momentum and real spaces. 
Such cyclic motions form Landau levels. However, at the edges of the top and bottom surfaces, such 
cyclic motions are not allowed, leading to skipping edge states.  
While the semiclassical analysis is based on ${\bm k}$ space for translational-invariant systems,
our results show the presence of the distinct profile of the local DOS for edge states on opposite 
surfaces in amorphous systems. 

\begin{figure}[t]
		\centering
	\includegraphics[width=0.75\textwidth]{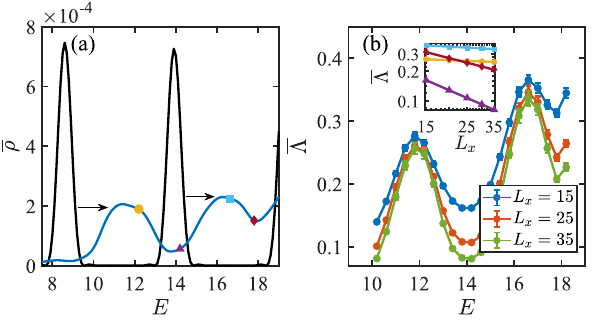}
	\caption{	
		(a) Sample averaged DOS $\overline{\rho}(E)$ versus the energy $E$ for amorphous systems (blue 
		line) and regular lattices (black line), computed by the KPM with the expansion order $N_c=2^{13}$.
		Here, $L_x=25$, $L_y=20$, and $L_z=44$.
		Arrows indicate the shifting direction of Landau levels induced by structural disorder.
		(b) Sample averaged normalized localization length $\bar{\Lambda}$ with respect to $E$ for 
		$L_x=15,25$ and $35$.   Here $L_y=20$ and $L_z=10^4$.
		The inset shows the finite-size scaling of $\bar{\Lambda}$ versus $L_x$ at $E=12.2$, $14.2$, $16.6$ and $17.8$
		[highlighted by yellow circle, purple triangle, blue square and red diamond in (a), respectively].
		For $E=14.2$ and $17.8$, we take $L_z=10^4$ and $L_y=20$, and
		for $E=12.2$ and $16.6$, we take $L_z=10^5$ and $L_y=20$ (the results are averaged over 40 samples).
	}		
	\label{Fig3}
\end{figure}

\section{Landau levels and localization properties}
To further elaborate on the mechanism of 3D quantum Hall
effect, we compute the DOS defined as
$\rho (E)=\sum_i \delta (E-E_i)/(2N)$, where $E_i$ is the $i$th eigenvalue of the Hamiltonian and $N=L_x L_y L_z$.
We use the kernel polynomial method (KPM) to compute the DOS with Chebyshev expansion up to
order $N_c$~\cite{Fehske2006RMP} under the same boundary conditions as for the Bott index.
To study the localization properties, we calculate the localization length of the system in a long bar 
geometry~\cite{Kramer1981PRL,Kramer1983ZPB} by
\begin{equation}\label{LL}
	\lambda^{-1}(E)=-\lim_{L_z\rightarrow \infty}\frac{1}{2L_z}\log \mathrm{Tr}\, [G_{1z}(E)G_{1z}^\dagger (E)],
\end{equation}
where $G_{1z}(E)=\langle 1|(E-H+i0^+)^{-1}|z \rangle $ is the submatrix of the Green's function between 
the site states in the first and final layer along $z$, which can be obtained iteratively after partitioning the bar into layers each 
with thickness $R_c$ along $z$~\cite{Mackinnon1985ZPB,Mackinnon1985ZPB2}.
In the calculations, we apply PBCs along $x$ and OBCs along $y$ and $z$.

Figure~\ref{Fig3}(a) displays the sample averaged DOS $\overline{\rho}(E)$ as a function of energy $E$ (blue line),
remarkably showing the existence of Landau levels despite the presence of strong structural disorder in amorphous
lattices. In stark contrast to the regular case (black line), the Landau levels are broadened, resulting in an overlap 
between neighboring levels. The system thus cannot form band insulators for the bulk states as in the 
regular case. Fortunately, these states are spatially localized as reflected by the decrease of the averaged normalized 
localization length $\overline{\Lambda}=\overline{\lambda}/L_x$ with the increasing of system sizes 
[see Fig.~\ref{Fig3}(b)]~\cite{Kramer1981PRL,Kramer1983ZPB}.
Since the localized states have no contribution to electrical transport, the quantized Hall conductance is
attributed to the edge states (see the following discussion).
Figure~\ref{Fig3}(b) also suggests that around the peaks of the two Landau levels, $\overline{\Lambda}$ 
converges to finite values as the system enlarges, signaling critical extended behaviors.
Such critical points correspond to the transition points between neighboring plateaus of the Hall conductance.
Specifically, we fit $\overline{\Lambda}$ computed for different system sizes as  
$\overline{\Lambda}\propto L_x^{-\alpha}$
at $E=12.2$ (yellow circle), $14.2$ (purple triangle), $16.6$ 
(blue square) and $17.8$ (red diamond), yielding $\alpha=0.07, 0.82, 0.09$ and $0.48$, 
respectively [see the inset of Fig.~\ref{Fig3}(b)].
The nonzero slope $\alpha$ suggests the localized character.
For $E=12.2$ and $16.6$, the slope $\alpha$ is very small, indicating their closeness to 
critical energies. 
Note that the critical energies are difficult to determine accurately in our 3D case due to the large
consumption of computational resources. 
This mechanism of the broadening of Landau levels and localization of bulk states is generic, implying that 3D quantum Hall effect can be observed as long as the Landau levels are not too close in the energy space, independent of the details of the model. 

Figure~\ref{Fig3}(a) also illustrates that the Landau levels in our amorphous lattices move to 
higher energy compared with its regular counterpart (black line), explaining the up-moving 
behavior of the Hall conductance plateau mentioned above .  

\begin{figure}
	\centering
	\includegraphics[width=0.95\textwidth]{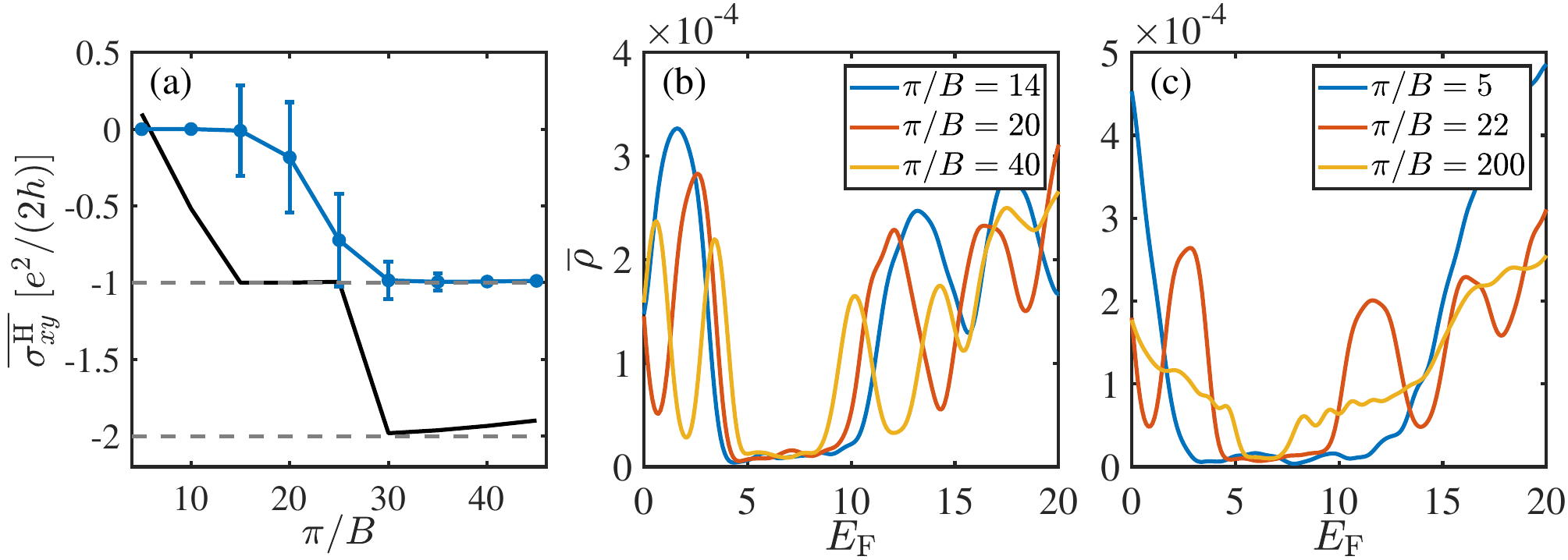}
	\caption{ (a) Sample averaged Hall conductance $\overline{\sigma_{xy}^{\text{H}}}$ 
			with respect to the inverse of the magnetic field $B$ at $E_\mathrm{F}=12$ for the regular lattice 
			(black line) with $L_x=300,L_y=20,L_z=40$ and amorphous lattices (blue line) with 
			$L_x=200,L_y=20,L_z=40$. (b-c) Sample averaged DOS calculated under various magnitudes of the magnetic field. 	
	}   	
	\label{Fig_magB}
\end{figure}

\section{Effect of the magnitude of the magnetic field}
To elaborate on the effect of the magnitude of the magnetic field on the 3D quantum Hall effect,
we calculate the sample averaged Hall conductance $\overline{\sigma_{xy}^{\text{H}}}$ with respect to the inverse of the magnetic field $B$ at a fixed $E_\mathrm{F}$, as shown in Fig.~\ref{Fig_magB}(a).
We see quantized plateaus in both regular (black line) and amorphous lattices (blue line) and 
as the magnitude of the magnetic field decreases ($\pi/B$ increases), the Hall conductance transitions from one plateau to another one with larger absolute value.
The transition is attributed to the shifting of the Landau levels to lower energy as the magnetic field decreases, as shown in Fig.~\ref{Fig_magB}(b) for amorphous lattices.  
We also observe that the plateaus of the Hall conductance on amorphous lattices float upwards compared with the regular lattice,
which is caused by the right moving of the Landau levels in the energy space induced by the structural disorder as we have explained in Fig.~\ref{Fig3}(a).  

The magnitude of the magnetic field cannot be too small
for the occurrence of 3D quantum Hall effect which can be understood easily by taking the limit 
of no magnetic field under which no quantum Hall effect arises.
When the magnetic field is too large, the 3D quantum Hall effect will also disappear.
Intuitively, as the strength of the magnetic field increases, the bulk states with energy further 
away from the Weyl points will be squeezed into Landau levels together with the Fermi arc states near the
Weyl points, because the degeneracy of the Landau level is proportional to the magnitude of the magnetic field.
Away from the Weyl points, the bulk DOS is large, reminiscent of the conventional 3D metal, leading to 
the disappearance of well-separated Landau levels even on the regular lattice.
Thus, the magnitude of the magnetic field has an upper bound for the 3D quantum Hall effect.
To justify our arguments above, we calculate the DOS on amorphous lattices as shown in Fig.~\ref{Fig_magB}(c) 
under magnetic fields with magnitude $\pi/B=5,22$ and $200$. 
Only for moderate magnitude $\pi/B=22$ (the red line), we see clear patterns of broadened Landau levels 
with stark contrast between peaks and valleys, which often implies the existence of quantum Hall effect.
Under too large (the blue line) and too tiny (the yellow line) magnetic field,
Landau levels cannot be identified, indicating the absence of quantum Hall effect.
Therefore, the occurrence of 3D quantum Hall effect requires the magnetic field in a moderate range of magnitude.

\begin{figure}
	\centering
	\includegraphics[width=0.75\textwidth]{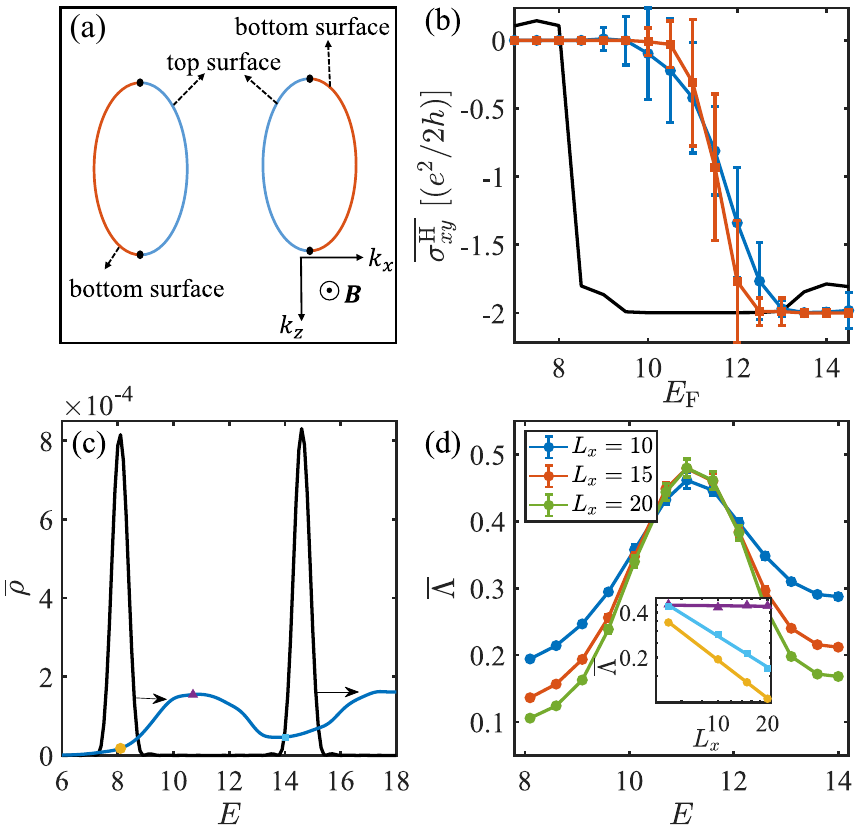}
	\caption{
		(a) Schematic of the Weyl points and Fermi arcs for the Hamiltonian (\ref{Ham2}) with TRS on regular lattices.
			(b) Configuration averaged Hall conductance versus the Fermi energy for
			the Hamiltonian (\ref{Ham2}) on amorphous (blue line) and regular (black line) lattices. 
			Th red line shows the Bott index on amorphous lattices.
			(c) Sample averaged DOS versus the energy $E$ for amorphous systems (blue 
			line) and regular lattices (black line).
			(d) Normalized localization length $\bar{\Lambda}$ averaged over 40 samples with respect to $E$ with $L_y=20$ and $L_z=10^4$.
			The inset shows the finite-size scaling of $\overline{\Lambda}$ versus $L_x$ at $E=8.1$, $10.7$ and $14$
			[highlighted by yellow circle, purple triangle and cyan square in (c), respectively]. 
	}
	\label{Fig4}
\end{figure}

\section{Model with TRS}
To demonstrate that 3D quantum Hall effect can be utilized to detect the
Weyl band like topological properties in amorphous systems with TRS,
we construct the following Hamiltonian,
\begin{equation}\label{Ham2}
	\hat{H}_T=\sum_{\bm r} [\hat{c}_{\bm r}^\dagger T_T^{0}\hat{c}_{\bm r}+
	\sum_{\bm d} t(|{\bm d}|)\hat{c}_{{\bm r}+{\bm d}}^\dagger 
	T_T(\hat{{\bm d}})\hat{c}_{\bm r}	]
\end{equation} 
with four internal degrees of freedom at each site, i.e. $\hat{c}_{\bm
	r}^\dagger=(\hat{c}_{{\bm r},1}^\dagger, \hat{c}_{{\bm r},2}^\dagger,\hat{c}_{{\bm 
		r},3}^\dagger,\hat{c}_{{\bm r},4}^\dagger)$.
Here
$T_T^0=M(k_w^2 \\ -6)\tau_z\sigma_0+(4D_2+2D_1)\tau_0\sigma_0+\beta \tau_y\sigma_y$ and 
$T_T(\hat{{\bm d}})=M\tau_z\sigma_0+
(i\gamma/2)(\hat{d}_x\tau_x\sigma_z+\hat{d}_y\tau_y\sigma_0)
-(D_2\hat{d}_x^2+\\ D_1\hat{d}_y^2+D_2\hat{d}_z^2)
\tau_0\sigma_0+i\alpha\hat{d}_y\tau_x\sigma_y/2$, with ${\tau_x}$ ($\nu=x,y,z$) being another
set of Pauli matrices referring to orbital degrees of freedom.
The Hamiltonian respects TRS, i.e. $\hat{T}\hat{H}_T\hat{T}^{-1}=\hat{H}_T$ 
with time-reversal operator $\hat{T}$ acting as $\hat{T}i\hat{T}^{-1}=-i$ and $\hat{T}\hat{c}_{\bm r}\hat{T}^{-1}
=i\tau_0\sigma_y\hat{c}_{\bm r}$.
In this section, we take $M=13, \alpha=-80, \beta=20, B=\pi/22$ and other parameters the same as 
before.

For a regular lattice, the Hamiltonian $\hat{H}_T$ describes a Weyl semimetal with double 
pairs of Weyl points and two pairs of Fermi arcs, as shown schematically in Fig.~\ref{Fig4}(a).
Without magnetic fields, the \textit{anomalous} Hall conductance contributed by each set of Fermi arcs 
cancels out each other. We thus cannot identify the topological property by probing the anomalous
Hall conductance. Fortunately, under magnetic fields along $y$, the two pairs of Fermi arcs both lead
to counterclockwise cyclic motions [see Appendix \ref{AppA}], which give rise to nonzero 3D
quantum Hall conductance [see the black line in Fig.~\ref{Fig4}(b) with system size $L_x=350$, $L_y=20$ and $L_z=40$]. 
In Fig.~\ref{Fig4}(b), we also plot the numerically 
computed Hall conductance with respect to the Fermi energy in amorphous lattices with $L_x=250$, $L_y=20$
and $L_z=40$ by the blue line, remarkably illustrating
the emergence of a quantized plateau at $\overline{\sigma_{xy}^\text{H}}=-2e^2/(2h)$ when 
$E_\text{F} \gtrsim 13$. Such a plateau arises from a transition from $\overline{\sigma_{xy}^\text{H}}=0$ to 
$\overline{\sigma_{xy}^\text{H}}=-2e^2/(2h)$ because the two pairs 
of Fermi arcs in our model are symmetric due to TRS and thus contribute identical 
Hall conductance.

In addition, we show the bulk-boundary correspondence by calculating 
the Bott index on amorphous lattices [see the red line in Fig.~\ref{Fig4}(b) with $L_x=25,L_y=20,L_z=44$], which is consistent with the Hall conductance. 
Similar to the case without time-reversal symmetry, the shift of the plateau of the Hall conductance in energy is attributed to the right-moving behavior of the Landau levels as shown in Fig.~\ref{Fig4}(c) for $L_x=40$, $L_y=20$ and $L_z=44$ with arrows indicating 
the shifting direction of Landau levels induced by structural disorder.
Although the Landau levels overlap with each other and renders the system gapless,
the bulk states are localized except at a critical point, demonstrated by the  normalized
localization length $\overline{\Lambda}$ in Fig.~\ref{Fig4}(d).
We see that $\overline{\Lambda}$ decreases as $L_x$ increases far away from $E=10.7$ and tends to converge to a constant near this energy.
The inset shows the result of the finite-size scaling calculations. For $E=8.1$ and $14$, by taking $L_z=10^4$, 
we observe a finite slope, indicating the localization behavior, whereas
for $E=10.7$, by taking $L_z=10^5$, we see a vanishing slope, which is a signature of the critical extended property.
Therefore, the quantized Hall conductance arises from the edge states that remain extended; meanwhile, the localized 
bulk states do not contribute to the transport. 
The generic mechanism indicates that 3D quantum Hall effect may be broadly used to detect the topology of 
time-reversal symmetric topological semimetals,
provided that the Landau levels are not too close in the energy space.  

\begin{figure}
	\centering
	\includegraphics[width=0.9\textwidth]{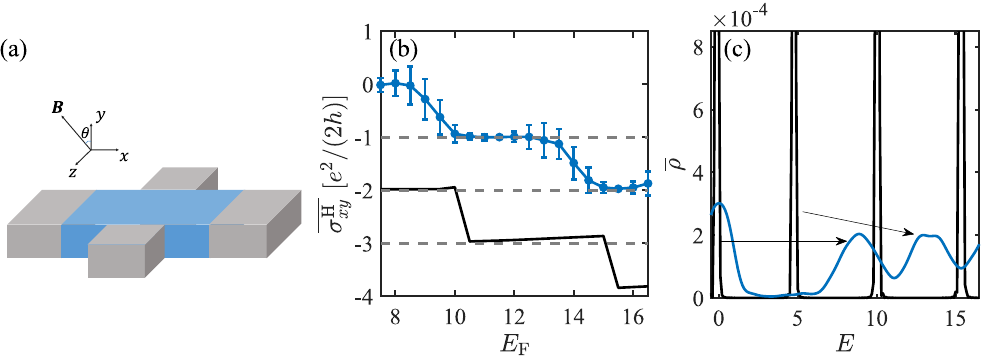}
	\caption{
		(a) Schematic of the sample under a tilted magnetic field ${\bm B}$.
		(b) Configuration averaged Hall conductance $\overline{\sigma^\text{H}_{xy}}$ 
		with respect to the Fermi energy $E_\text{F}$
		for the Hamiltonian (\ref{Ham1}) on amorphous 
		(the blue line) and regular (the black line) lattices.
		The sample size is $L_x=200,L_y=20,L_z=40$. 
			(c) The DOS for regular (black line) and amorphous (blue line) lattices for $L_x=25,L_y=20,L_z=44$ 
			with shifting of Landau levels induced by structural disorder indicated by arrows.   }
	\label{SM1}
\end{figure}
\begin{figure}
	\centering
	\includegraphics[width=0.9\textwidth]{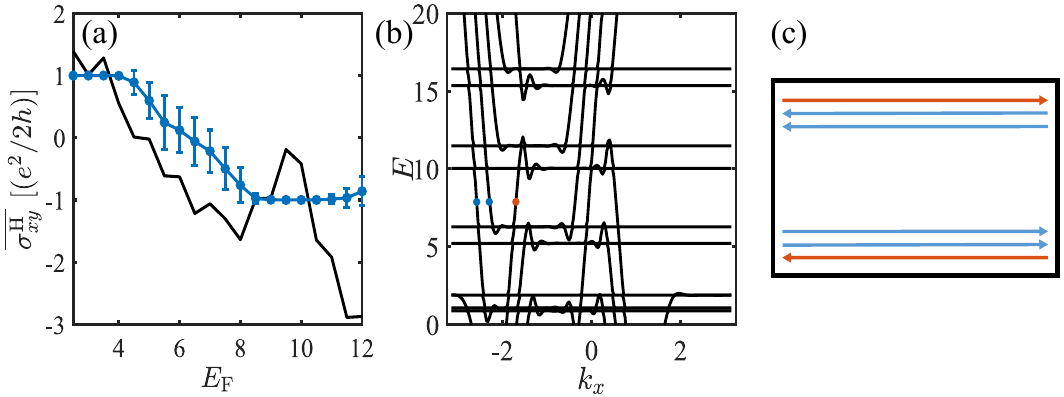}
	\caption{	
		(a) Sample averaged Hall conductance of the time-reversal invariant Hamiltonian (\ref{Ham2}) under a magnetic field with $B_y=\pi/22$ and $\theta=10^\circ$. 
		The system size for amorphous lattice (blue line) is $L_x=250,L_y=20,L_z=40$ and for regular
		lattice (black line) is  $L_x=550,L_y=20,L_z=50$.
		(b) The Landau level spectrum on a regular lattice versus $k_x$ with PBCs in the $x$ direction. 
		Here $L_y=20$ and $L_z=60$ with OBCs along both directions.
		At $E=8$, the states at one edge are indicated by blue and red points. 
		(c) Schematic of edge states projected onto a single plane with arrows indicating 
		the propagating directions of the electrons.  	}
	\label{SM3} 
\end{figure}
\section{Tilted magnetic field}
In this section, we will show that it also occurs under a tilted magnetic 
field. Specifically, 
we study the Hamiltonian $\hat{H}$ (\ref{Ham1}) under a magnetic field 
$\bm{B}=(B_x,B_y,B_z)$ lying in the $(y,z)$ plane with a tilting angle $\theta$ from the $y$ axis,  
i.e. $B_x=0$ and $B_z=B_y\tan \theta$, as shown in Fig.~\ref{SM1}(a).
Here, we take $B_y=\pi/22$ and $\theta=10^\circ$ and the calculated Hall conductance for amorphous 
(the blue line) and regular (the black line) lattices is displayed in Fig.~\ref{SM1}(b).
We see that the Hall conductance on amorphous lattices is quantized at 
$\overline{\sigma^\text{H}_{xy}}=-e^2/(2h)$
for $10 \lesssim E_\text{F} \lesssim 13.5$ and at $\overline{\sigma^\text{H}_{xy}}=-2e^2/(2h)$ for
$15 \lesssim E_\text{F} \lesssim 16$, showing the existence of 3D quantum Hall effect on 
amorphous lattices.
The Hall conductance plateau at a certain value of $\overline{\sigma^\text{H}_{xy}}$ moves to
higher 
Fermi energy $E_\text{F}$ compared to
the Hall conductance on a regular lattice, attributed to the shifting of Landau levels to the higher energy 
as shown in the DOS in Fig.~\ref{SM1}(c)
(in amorphous lattices, the Landau levels shift roughly the distance of two Landau levels 
compared with the regular lattice, indicated by the arrows). 
We also observe that although the component of the magnetic field perpendicular to the top and bottom 
surface, namely $B_y$ is the same as in Fig.~\ref{Fig2}, the Hall conductance is different at the same 
Fermi energy $E_\text{F}$, implying the angle 
dependence of 3D quantum Hall effect on amorphous lattices, similar to the regular case~\cite{Xie2020PRL}.  
Note that the Hall conductance is not well quantized at
$\overline{\sigma^\text{H}_{xy}}=-2e^2/(2h)$ 
for $10.5 \lesssim E_\text{F} \lesssim 15$
and at $\overline{\sigma^\text{H}_{xy}}=-4e^2/(2h)$ for 
$15.5 \lesssim E_\text{F} \lesssim 16.5$ on the
regular lattice due to finite-size effect. 

We also study the time-reversal invariant Hamiltonian (\ref{Ham2})  under a tilted magnetic field with $\theta=10^\circ$.
	Surprisingly, on amorphous lattices we see a transition from a plateau of Hall conductance at $\overline{\sigma^\text{H}_{xy}}=e^2/(2h)$ to a plateau at $\overline{\sigma^\text{H}_{xy}}=-e^2/(2h)$ as shown in Fig.~\ref{SM3} (the blue line),
	in stark contrast to the case under a vertical magnetic field with Hall conductance quantized at even 
	integer values in units of $e^2/(2h)$, shown in Fig.~\ref{Fig4}(b).
	However, on a regular lattice no quantized plateaus of Hall conductance arise in the same energy region up to $L_x=550$ and
	no signature of quantization is observed during the increasing of the system size.
	To elaborate on the reason why quantized Hall conductance emerge on amorphous lattices while is
	absent on the regular lattice,
	we show the energy spectrum with respect to the lattice momentum $k_x$ on a regular lattice
	in Fig.~\ref{SM3}(b), exhibiting Landau levels with edge states going upwards 
	and downwards alternately.
	For example, within the gap near $E=8$ (indicated by the blue and red points), there exist two edge states 
	propagating in the same direction (indicated by blue points) and one in opposite direction (indicated 
	by the red point) on each edge, also shown schematically in Fig.~\ref{SM3}(c). 
	The quantized Hall conductance is prohibited by the backscattering between the counter-propagating
	edge states on each edge.
	On amorphous lattices, a pair of counter-propagating edge states become localized due to disorder 
	with only one edge channel left, resulting in the quantized Hall conductance.

\section{Conclusion}
In summary, we have theoretically demonstrated the existence of 3D quantum Hall effect in a slab of
topological amorphous metals subjected to magnetic fields. We find that while structural disorder
significantly broadens the Landau levels such that two neighboring Landau levels overlap with
each other, the bulk states are spatially localized except at critical transition energies so that they
do not contribute to the Hall conductance. However, due to the topological property of the
Landau levels revealed by the Bott index, the edge states arise at distinct hinges on two 
opposite surfaces, leading to 3D quantum Hall effect. 
The mechanism of broadening of Landau levels and localization of bulk states is quite generic, independent of the model details.
Our results indicate that Weyl band like topology in amorphous materials can be identified
through measuring the 3D quantum Hall effect. 
We also demonstrate the existence of 3D quantum Hall effects 
in the amorphous system under a tilted magnetic field. 
Our results suggest that 3D quantum Hall effects may broadly arise
in amorphous Weyl like materials.
In fact, 3D quantum Hall effect has been experimentally observed in crystalline topological semimetal
materials $\mbox{Cd}_3\mbox{As}_2$~\cite{Xiu2017NC2,Kawasaki2017NC,Stemmer2018PRL,Xiu2019Nature},
which will stimulate further experimental study in broader contexts, including non-crystalline systems. 
 To find concrete material realization of 3D quantum Hall effect in amorphous systems is an important and 
 promising direction for further research.
In addition, our results demonstrate the presence of Landau levels in 3D amorphous systems, which
may stimulate further investigation into the quantum Hall effect in amorphous topological insulators~\cite{Hellman2023NM}, 
such as $\text{Bi}_2 \text{Se}_3$, under magnetic fields.

\section*{Acknowledgements}
We thank Y.-B. Yang and B. Wieder for helpful discussions. 
We also acknowledge the support by Center of High Performance
Computing, Tsinghua University.


\paragraph{Funding information}
The work is supported by the National Natural Science Foundation
of China (Grant No. 11974201 and 12474265), Tsinghua University Dushi Program
and Innovation Program for Quantum Science and Technology (Grant No. 2021ZD0301604).

\begin{appendix}

\section{Surface spectral function without magnetic fields}
\label{AppB}
In this Appendix, we will calculate the surface spectral function of Hamiltonian (\ref{Ham1}) 
under
no magnetic field to show ARPES measurement may be difficult for topological amorphous metals.

\begin{figure}
	\centering
	\includegraphics[width=0.75\textwidth]{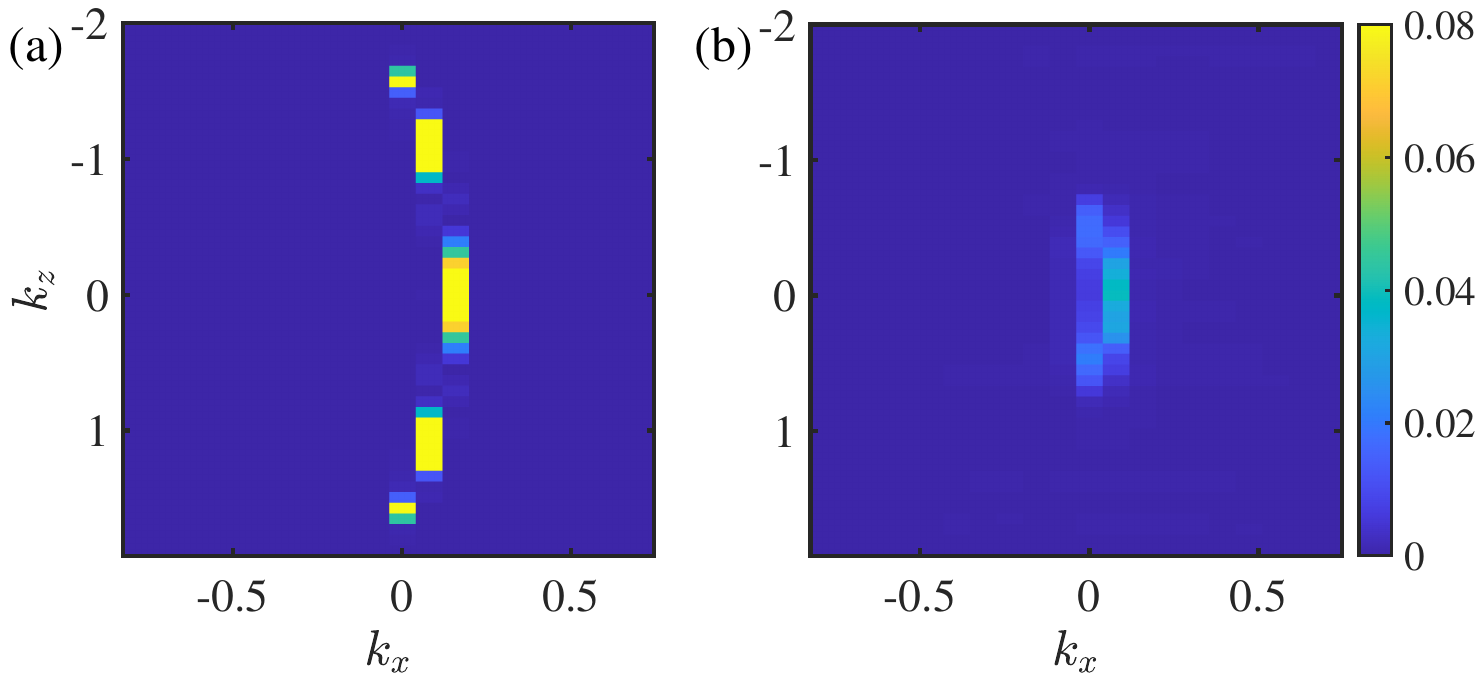}
	\caption{Surface spectral function for the top surface at $E=9$ on regular (a) and amorphous (b) lattices.} 
	\label{SM4}
\end{figure}

The observable extracted by ARPES is the surface spectral function
 \begin{equation}\label{Aspec}
 A(\bm{k},E)=\sum_l \langle \bm{k},l|\delta(H-E)|\bm{k},l\rangle,
 \end{equation} 
where $|\bm{k},l\rangle$ is the plane wave with wave vector $\bm{k}=(k_x,k_z)$ restricted in the surface layer 
in the $y$ direction associated with internal degree of freedom $l$.
Here we take $H$ as the Hamiltonian (\ref{Ham1}) under no magnetic field and $E$ as the energy of the Weyl point.
On the regular lattice, we see a clear Fermi arc on the top surface, as shown in Fig.~\ref{SM4}(a).
Note that the discontinuity of the Fermi arc is due to the finite size of the system.
On the other hand, on amorphous lattices as shown in Fig.~\ref{SM4}(b), the signature of the Fermi arc is much more vague, 
suggesting the difficulty in detecting the topology by ARPES here.

\section{Explanation of the constraint on the thickness along the $y$ direction }
\label{AppC}
In the main text, we use a slab geometry thin along $y$ 
for the calculation of the Hall conductance because a finite thickness along $y$ is 
necessary to the 3D quantum Hall effect even in the crystalline Weyl semimetal. 
In this Appendix, we will explain the reason in detail.

\begin{figure}[t]
 	\centering
 	\includegraphics[width=\textwidth]{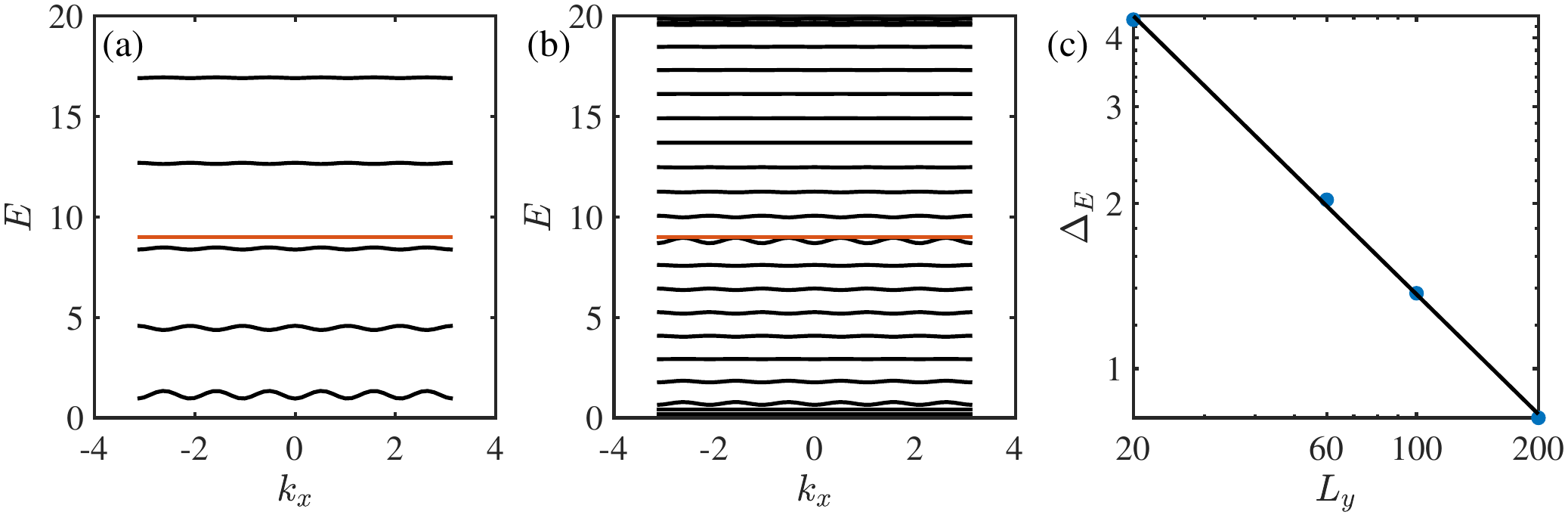}
 	\caption{(a)-(b) Landau level spectra of Hamiltonian (\ref{Ham1}) on a regular lattice 
 			under a magnetic field $B=\pi/22$ near the energy of the Weyl point (the red line). $L_y=20,L_z=44$ for (a) and $L_y=60,L_z=44$ for (b).
 	(c) Finite-size scaling of the energy gap $\Delta_E$ with respect to $L_y$ at $L_z=44$. }
 	\label{SM5}
 \end{figure}

We show the energy spectrum of Hamiltonian (\ref{Ham1}) under a magnetic field with $B=\pi/22$ 
in Fig.~\ref{SM5}(a)-(b), with the energy of the Weyl point indicated by the red line. 
Here we consider a regular lattice with PBCs in both $x$ and $z$ directions.
Note that the Landau levels are not perfectly flat due to the lattice effect.
We take $L_y=20$ for Fig.~\ref{SM5}(a), $L_y=60$ for Fig.~\ref{SM5}(b) and $L_z=44$ for both.
Obviously, the Landau levels for $L_y=60$ are much denser than $L_y=20$, indicating the
 decrease of separation between Landau levels as the thickness in the $y$ direction increases.
We show the energy separation between Landau levels $\Delta_E$ as a function of $L_y$ in logarithmic scale in Fig.~\ref{SM5}(c), 
where $\Delta_E$ is defined as the magnitude of the energy gap within which the energy of the Weyl point lie at $k_x=0$.
We obtain $\Delta_E \propto L_y^{-0.72}$, indicating that the Landau levels will get closer and closer in the energy 
space with the increasing of the thickness of the sample and touch each other in the thermodynamic limit, leading to the disappearance of the 3D quantum Hall effect.
Therefore, the 3D quantum Hall effect needs a finite thickness along $y$ for a sample even in the crystalline case.
Similarly, in the presence of disorder, the Landau levels cannot be too close in energy for the quantum Hall effect, 
so the sample cannot be too thick.

\section{Semiclassical explanation of the 3D
	quantum Hall effect for a time-reversal invariant Hamiltonian on regular lattices}
	\label{AppA}
In the main text, we have shown that 3D quantum Hall effect can arise in a time-reversal 
invariant Hamiltonian $\hat{H}_T$ with double pairs of Weyl points on regular lattices under a
magnetic field.
In this Appendix, through semiclassical analysis based on the semiclassical equations of motion~\cite{Xie2020PRL}, 
we will account for why the two pairs of Fermi arcs give rise to nonzero 
quantum Hall conductance rather than cancel out the contribution from each other as for the
{\it anomalous} Hall conductivity without magnetic fields.

\begin{figure}[t]
	\centering
\includegraphics[width=0.75\textwidth]{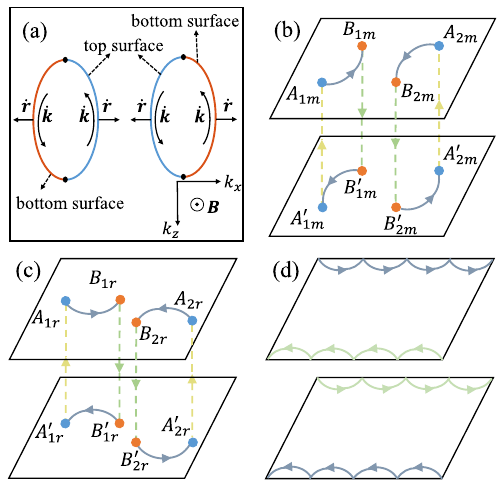}
	\caption{
		(a) Schematic of Weyl points and Fermi arcs on top and bottom surfaces projected onto a single 
		plane for the Hamiltonian $\hat{H}_T$ on a regular lattice in the main text. 
		The velocities of the center of mass of an electronic wave packet $\dot{\bm k}$ and $\dot{\bm r}$ in
		momentum and real space, respectively, under a magnetic field in $y$ direction are indicated in the
		figure.
		(b) Schematic of cyclic motions of electrons in momentum space where Weyl points are denoted
		by
		$A_{im}$ ($A^\prime_{im}$) and $B_{im}$ ($B^\prime_{im}$) with $i=1,2$.
		The chiral Landau levels tunneling from top to bottom surface and vise versa are colored green and 
		yellow, respectively.
		(c)  Schematic of cyclic motions of electrons in real space.
		(d) Schematic of edge states contributed by the left pair (colored grey) and right pair
		(colored green) of Fermi arcs.
	}
	\label{SM2}
\end{figure}

The double pairs of Weyl points and two loops of Fermi arcs on top and bottom surfaces of the 
Hamiltonian in a regular lattice 
are schematically shown in Fig.~\ref{SM2}(a).
For clarity, we project the Fermi arcs on both surfaces onto the same plane.
At Weyl points, Fermi arcs on opposite surfaces are connected by chiral Landau levels.
The semiclassical equation of motion 
$\dot{\bm r}=\partial \varepsilon _{\bm k}/(\hbar \partial {\bm k})$ tells us that the real space velocity
$\dot{\bm r}$ is perpendicular to the Fermi arc.
In our case, $\dot{\bm r}$ points out of the loops formed by Fermi arcs, as indicated in 
Fig.~\ref{SM2}(a).
Under a magnetic field along $y$, the electronic wave packets cycle along the loops of 
Fermi arcs counterclockwise in momentum space, also shown in Fig.~\ref{SM2}(a), determined by
the equation 
$\hbar \dot{\bm k}=-e\dot{\bm r}\times {\bm B}$.
Specifically, as illustrated in Fig.~\ref{SM2}(b), for an electron starting from the Weyl point
$A_{1m}$ on the top surface in momentum
space, it will move along the Fermi arc to another Weyl point $B_{1m}$ and then tunnel to
$B^\prime_{1m}$ on the bottom surface.
Later, the electron travels along the Fermi arc on the bottom surface to $A^\prime_{1m}$ and tunnel 
back
to $A_{1m}$, completing a cyclic motion in momentum space. 
Correspondingly, in real space, the electron moves from $A_{1r}$ to $B_{1r}$ and then tunnel to
$B^\prime_{1r}$ on the bottom surface, as shown in Fig.~\ref{SM2}(c).
After traveling to $A^\prime_{1r}$ and tunneling back to $A_{1r}$, a cyclic motion in real space is 
completed.
The cyclic behavior of electrons on the other pair of Fermi arcs
[between $A_{2m}$ ($A^\prime_{2m}$) and $B_{2m}$ ($B^\prime_{2m}$)] and corresponding
real space motions can be extracted in a similar manner.
Viewed from the positive direction of the $y$ axis, in both momentum and real space, 
the cyclic directions of electrons dominated by two 
loops of Fermi arcs are both counterclockwise. 
In fact, the same cycling direction for different pairs of Fermi arcs is ensured by the 
time-reversal symmetry.
At the edge, the cyclic motions cannot be finished, giving rise to edge states, as shown in 
Fig.~\ref{SM2}(d).
The edge states contributed by the left pair of Fermi arcs are indicated by grey color and those 
contributed by the right
pair are colored green.
Under the setup of Hall conductance measurement in Fig.~\ref{Fig2}(b) in the main text, the two pairs of edge 
states contribute quantum Hall conductance with the same sign and nonzero total quantum Hall 
conductance is measurable.

\end{appendix}




\end{document}